# Ultra-thin Epitaxial MgB$_2$ on SiC: Substrate Surface Polarity Dependent Properties

Weibing Yang[1][†][*], Leila Kasaei[2][†], Hussein Hijazi[2], Sylvie Rangan[2], Yao-wen Yeh[2], Raj K Sah[1], Jay R Paudel[1], Ke Chen[1], Alexander X. Gray[1], Philip Batson[2], Leonard C. Feldman[2], and Xiaoxing Xi[1]

[1]*Department of Physics, Temple University, Philadelphia, PA 19122, USA*

[2]*Department of Physics and Astronomy and Laboratory for Surface Modification, Rutgers University, New Brunswick, NJ 08901, USA*

*Corresponding authors

†These authors contributed equally to this work

**Abstract:** High quality, ultrathin, superconducting films are required for advanced devices such as hot-electron bolometers, superconducting nanowire single photon detectors, and quantum applications. Using Hybrid Physical-Chemical Vapor Deposition (HPCVD), we show that MgB$_2$ films as thin as 4 nm can be fabricated on the carbon terminated 6H-SiC (0001) surface with a superconducting transition temperature above 33K and a rms roughness of 0.7 nm. Remarkably, the film quality is a function of the SiC surface termination, with the C-terminated surface preferred to the Si-terminated surface. To understand the MgB$_2$ thin film/ SiC substrate interactions giving rise to this difference, we characterized the interfacial structures using Rutherford backscattering spectroscopy/channeling, electron energy loss spectroscopy, and x-ray photoemission spectroscopy. The MgB$_2$/SiC interface structure is complex and different for the two terminations. Both terminations incorporate substantial unintentional oxide layers influencing MgB$_2$ growth and morphology, but with different extent, intermixing and interface chemistry. In this paper, we report measurements of transport, resistivity, and critical superconducting temperature of MgB$_2$/SiC that are different for the two terminations, and link interfacial structure variations to observed differences. The result shows that the C face of SiC is a preferred substrate for the deposition of ultrathin superconducting MgB$_2$ films.

1. Introduction

Ultra-thin superconducting films of MgB$_2$ have attracted much interest [1-4] owing to their relatively high $T_c$ of 39 K, large coherence lengths, and high critical current density [5,6] for applications in bolometers, photon detectors, and quantum devices [7-9]. High quality MgB$_2$ films have been fabricated successfully using Hybrid Physical-Chemical Vapor Deposition (HPCVD) [10], creating thin films with high $T_c$ and low residual resistivity.

However, the Volmer–Weber like island growth mode in the HPCVD process is detrimental to the production of ultrathin and smooth $MgB_2$ films [11-14], where coalescence of islands often leads to rough surfaces. For many superconducting electronic applications ultrathin (~5 nm), smooth (ideally with RMS roughness < 1 nm) and uniform films are required. Previously, we reported on low-angle ion milling to thin 40 nm HPCVD $MgB_2$ films and fabricated 5 nm superconducting films with $Tc$ as high as 36 K [2]. Novoselov et al. have reported growth of 10 nm HPCVD $MgB_2$ films directly on the Si terminated SiC (0001) substrate with $T_c$ as high as 35 K [15]. Recently, we have found that ultrathin $MgB_2$ films grown on C-terminated 6H-SiC substrate (0001bar) are significantly smoother than those on Si-terminated substrates and possess high quality electronic properties and high $Tc$ [16].

Interfacial phenomena are critically important in the design and manipulation of thin film functional materials[17-20]. Polarity of substrate is one of the important paramaters determining the structural, electric and magnetic properties of materials grown on polar materials. The polarity of SiC has been proved to have significant influence on the growth as well as properties of thin films such as graphene, GaN and AlN thin films [21-23]. Here, we report a comprehensive characterization of ultrathin HPCVD $MgB_2$ films grown on SiC substrates with both Si- and C-termination. The interface characteristics are correlated with measurements of $MgB_2$ transport, resistivity, and critical superconducting temperature comparing growth on these two principal SiC faces. A significant finding is the existence of a substantial magnesium oxide layer at the $MgB_2$/SiC interface with thickness and roughness dependent on the termination of the SiC substrate. The $MgO_x$ layer is thinner and smoother on the C-face than on the Si-face. The smoother $MgO_x$ layer leads to a smoother $MgB_2$ ultrathin film on the C-face. Overall, the result shows that the C face is a preferred substrate for the deposition of ultrathin superconducting $MgB_2$ films on SiC. To our knowledge these $MgB_2$ films are at the leading edge of this technology, combining the requirements of high $Tc$, thickness, and uniformity for advanced applications. The interface characterization reported here provides details and possible explanations for the growth habit and suggests procedures broadly applicable to superconducting thin film growth.

## II. Experimental details

SiC substrates used were the 6H polytype in the (0001) or (0001bar) direction. As described in Ref. [16], a double-side polished 6H-SiC(0001) substrate is C-terminated on one side and Si-terminated on the other, leading to surfaces of different polarities. Details of the HPCVD method for growing $MgB_2$ thin films have been described previously [10]. The HPCVD growth condition has been optimized to minimize the RMS roughness (a flow rate of 10 sccm diborane gas mixture, 5% $B_2H_6$ in $H_2$, was used in this work as compared to 2 sccm in our previous work [16]), (see section 1 of Supplementary Material [24]). Transport properties of $MgB_2$ films were characterized by standard four-point measurements [25] where probes were placed in the four corners of 5×5 $mm^2$ square sample. Resistivity vs. temperature measurement was carried out by dipping the sample into a liquid helium dewar immediately after removal from the HPCVD system to minimize air exposure. The measured resistance is converted into resistivity based on the van der Pauw solution for the square shape sample [26].

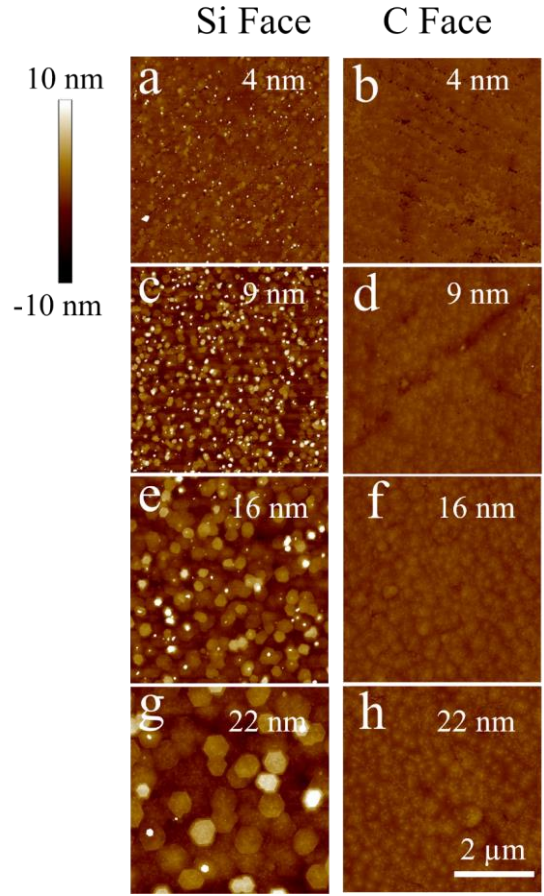

Figure 1. (a),(c), (e) and (g) AFM images for MgB$_2$ films grown on Si-terminated SiC with thicknesses of 4 nm, 9 nm, 16 nm and 22 nm respectively; (b), (d), (f) and (h) are the corresponding MgB$_2$ films on the C-terminated SiC that were grown at same conditions as (a), (c), (e) and (g).

Atomic force microscopy (AFM), scanning transmission electron microscopy (STEM), x-ray photoelectron spectroscopy (XPS), and Rutherford backscattering spectrometry (RBS) were used to characterize the film and the interface. AFM measurements were performed using a Veeco atomic force microscope. Imaging and electron energy loss spectroscopy were carried out using the Rutgers Nion UltraSTEM microscope operated at 60 kV with the convergence and collection semi-angles at 35 and 16.5 mrad, respectively. XPS measurements were performed in a Thermo K-alpha system with charge compensation using Al-Kα radiation and overall energy resolution of 0.7 eV. Under these conditions, the surface hydrocarbons were found at a binding energy of 284.7 eV. RBS measurements were performed using a General Ionex tandem accelerator with 1.6 MeV He$^{++}$ ions and a scattering angle of 130°. The sample was held approximately normal to the ion beam. The estimated depth resolution for Mg is ~ 25 nm. Ion beam channeling was performed along the (0001) direction of the 6H-SiC substrate and non-channeling (random) spectra were acquired by randomly varying incident angles. SIMNRA 7.01 software was used for non-channeling RBS spectra analysis [27].

**III. Results**

Table 1. RMS roughness, $T_{c0}$, residual resistivity $\rho_0$, and $\Delta\rho$ of MgB$_2$ films on C-SiC and Si-SiC.

|  | Film Thickness (nm) | 4 | 9 | 16 | 22 |
|---|---|---|---|---|---|
| C face | RMS (nm) | 0.7 | 0.6 | 0.5 | 0.7 |
|  | $T_{c0}$ (K) | 33.6 | 37.5 | 39.4 | 39.8 |
|  | $\rho_0$ ($\mu\Omega\cdot$cm) | 14.9 | 4.9 | 1.9 | 1.3 |
|  | $\Delta\rho$ ($\mu\Omega\cdot$cm) | 14.2 | 10.6 | 8.3 | 7.5 |
| Si face | RMS (nm) | 1.7 | 3.2 | 2.9 | 2.7 |
|  | $T_{c0}$ (K) | 35.4 | 37.8 | 39.2 | 40.6 |
|  | $\rho_0$ ($\mu\Omega\cdot$cm) | 24.7 | 5.3 | 2.3 | 1.4 |
|  | $\Delta\rho$ ($\mu\Omega\cdot$cm) | 24.3 | 10.6 | 8.1 | 7.7 |

### i. Deposition rates and surface roughness

As described in detail in the supplementary information [24] the basic growth method involves liquified Mg, combined with a flow of diborane gas to form MgB$_2$. In previous work [16], we used flow rates of 1 sccm and 2 sccm diborane gas mixture in the HPCVD deposition and obtained an RMS roughness of 1.2 nm and $T_c$ of 34.3 K in a 5.7 nm MgB$_2$ thin film on the C face of the SiC substrate. From a more recent systematic optimization of the diborane gas mixture flow rate, we found that 10 sccm produces the smallest roughness with the best superconducting properties. Figure 1 shows AFM images of MgB$_2$ thin films with nominal thicknesses of 4 nm, 9 nm, 16 nm, and 22 nm grown on the Si and C faces. Films with the same thickness but different terminations were grown in the same deposition run to ensure identical growth conditions. For all thicknesses, films on the C face are smoother than those on the Si face. Films on the Si face show clusters that are absent on the films on the C face. Energy dispersive x-ray spectroscopy (EDS) analysis shows that the clusters are MgB$_2$ grains. In addition, the films on the Si face show taller islands that are not completely connected while films on the C face show much better connectivity. On the Si face, the MgB$_2$ islands become larger for thicker depositions, typical for the island growth. Films of the same nominal thickness on the C face show a much smoother surface and don't have typical hexagonal MgB$_2$ grain as the case on Si face. The RMS roughness for the MgB$_2$ films in Fig. 1 is summarized in Table 1 showing values from 0.5 nm – 0.7 nm for the C face and 2 – 3 nm for the Si face. The result is a marked improvement from those in Ref. [16] and indicates that MgB$_2$ films grown on the C can be smoother than those on the Si face. Films composed of grains with a size comparable to the total film thickness are intuitively rougher.

### ii. Electronic properties

Figure 2 shows corresponding resistivity vs. temperature ($\rho$-$T$) curves for the MgB$_2$ films in Fig. 1. As the film thickness decreases, $T_c$ decreases and the residual resistivity $\rho_0$ increases for films on both the Si and C faces. The $T_c$ of MgB$_2$ films on the Si-face is slightly higher than the films on C-face, which is probably due to the biaxial tensile strain between MgB$_2$ grains as the MgB$_2$ growth mode in Si-face is more like island growth mode compared to the C face [28]. The results are summarized in Table 1. While the values of $T_c$ are similar to our earlier report [16], the $\rho_0$ values of the current films are much lower. The results are similar for both substrate terminations except for the thinnest films. The residual resistivity is much higher for the 4 nm film on the Si face than on the C face.

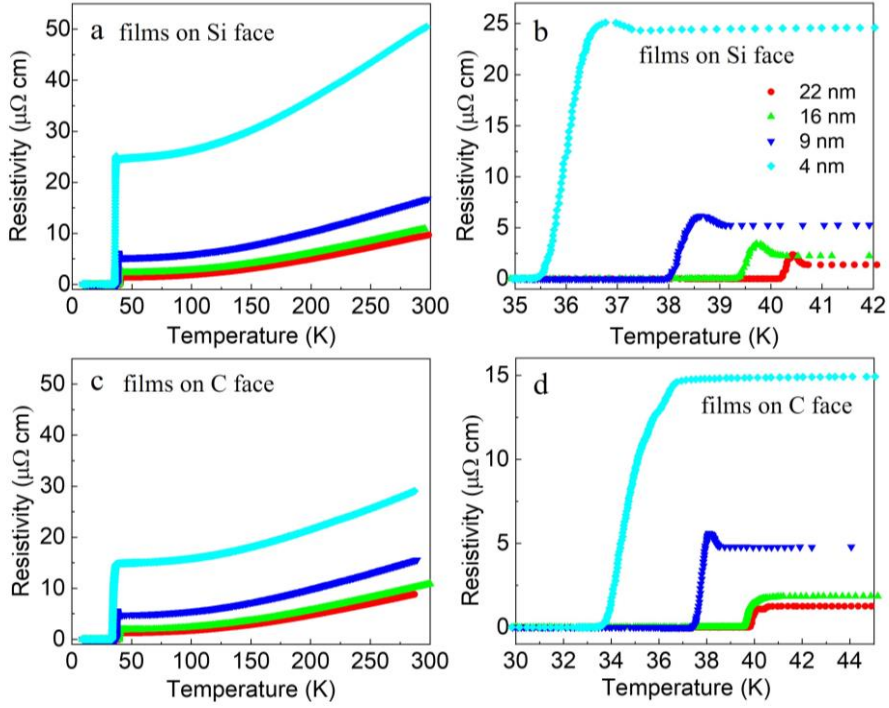

Figure 2. Resistivity vs. temperature curves for MgB$_2$ films on the Si face (a and b) and the C face (c and d).

Also included in Table 1 is $\Delta\rho$, the change in resistivity from room temperature to just above the superconducting transition temperature. Rowell has shown that this quantity, the room temperature, and residual resistivity difference can be used to quantify the grain connectivity in MgB$_2$ samples [29], with larger $\Delta\rho$ indicating poorer connectivity. The dependence of $\Delta\rho$ on film thickness is shown in Fig. 3a. At 22 nm, the films on both Si and C faces show $\Delta\rho$ values similar to our thicker, high quality MgB$_2$ films, indicating excellent grain connectivity. As the nominal thickness of the ultrathin film decreases, the grain connectivity effect becomes more and more important, reflected as a gradual increase in $\Delta\rho$. At 4 nm, the connectivity degrades rapidly, leading to a large $\Delta\rho$ increase for both faces. The film on the Si face shows much poorer connectivity than that on the C face. The conclusion on the grain connectivity is corroborated by the residual resistivity data. While $\Delta\rho$ reflects the temperature dependence of the electron-phonon scattering and grain connectivity, $\rho_0$ is determined by the grain connectivity and scattering of electrons by impurities, defects, as well as surfaces [29]. We have shown previously [30] that for clean MgB$_2$ films fabricated by HPCVD, the mean free path of electron scattering is limited by the film thickness. For example, changing the film thickness from 22 nm to 4 nm results in a decrease in the mean free path and thus an increase in electron scattering by a factor of 5.5. Combined with a reduction of grain connectivity, deduced from the increase in $\Delta\rho$, by a factor of 1.9 in the case of films on the C face, one can predict the $\rho_0$ value of the 4 nm film from that of the 22 nm film. The result, $\rho_0 = 1.3 \times 5.5 \times 1.9 = 13.7$ μΩ·cm for the 4 nm film, agrees well with the experimentally measured value. The results of the same procedure for all the films studied are shown in Figs. 3(b) and 3(c).

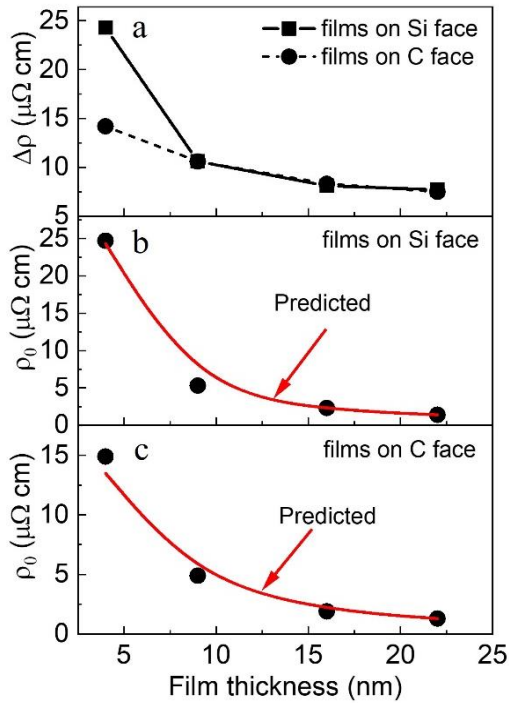

Figure 3. The change of resistivity from 300 K to 40 K and residual resistivity as function of film thickness on both Si and C face. The red curves are values expected based on the reductions of the electron mean free path and the grain connectivity as the film thickness decreases.

### iii. Interface characterization

To understand the influence of the SiC surface termination on the properties of the ultrathin $MgB_2$ films, we investigated the interfacial structure and chemistry of the $MgB_2$/SiC interface. The $MgB_2$ samples for interface characterizations are prepared separately using the same growth conditions as films described above to ensure the properties of $MgB_2$ films are consistent throughout this work.

#### iii-a) RBS channeling and interfacial oxides

RBS/channeling measurements were performed to provide depth-dependent information on both the composition ('random spectra') and crystallinity. Channeling, the reduction of scattering yield when the beam is aligned with a major crystallographic direction, yields information on crystal quality and identifies the alignment and composition of buried layers. The results show that for both SiC terminations, there is clearly an interfacial layer between the $MgB_2$ film and the SiC substrate containing both magnesium and oxygen (see section 2 of Supplementary Material [24]). The result of a composition analysis at the interface for the two samples, identified as $MgB_2/MgO_x$/SiC, is presented in Table 2, where the interfacial oxygen is ascribed to a $MgO_x$ layer. Within the experimental error, the composition of the interfacial layer is close to MgO. The thickness of the $MgO_x$ layer is estimated to be ~

Table 2. Atomic composition of MgB$_2$/SiC determined by RBS

| MgB$_2$/MgO$_x$/SiC | O$_{interface}$/cm$^2$ ($\times 10^{15}$) | Mg$_{interface}$/cm$^2$ ($\times 10^{15}$) | Thickness of MgO$_x$ interface(nm) |
|---|---|---|---|
| Si-Face (Ch-RBS) | 13 ± 2 | 12 ± 2 | 2.3 ± 0.2 |
| C-Face (Ch-RBS) | 6 ± 2 | 4 ± 2 | 0.9 ± 0.2 |

2.3 nm for the Si termination and ~ 0.9 nm for the C termination. Furthermore, there is a consistent, but small, channeling effect in the MgO$_x$ itself, indicating that the oxide is crystalline. We suggest the Mg-interface peaks result from a partially crystalline MgO layer at the interface, with orientation affected by the lattice mismatch between MgO and SiC, and then influenced by the mismatch with overlayer MgB$_2$ and MgO$_x$. (RBS also detected monolayer scale surface impurities of silicon and carbon at the surface of the MgB$_2$ overlayer that play no apparent role in the interface formation, but are noted here for completeness). The observation of a substantial oxide interfacial layer is a major new finding. The "buried oxide" is shown to be consistent with magnesium oxide by the chemical shift as observed in angular dependent high energy XPS (HAXPES) analysis (section 3 of Supplementary Material [24], also see the Ref. [31-37] therein).

The observation of a MgO$_x$ interfacial layer raises two questions: 1) the origin of the oxygen since the HPCVD process, being entirely conducted in a reducing environment, in principle eliminates the oxygen from the film growth? and 2) the roles of the oxide layer in determining the properties of the ultrathin MgB$_2$ films associated with different terminations? To address these questions, we investigated the first stages of the HPCVD process itself by heating the substrate along with the Mg pieces, without the introduction of the B$_2$H$_6$ gas mixture. Specifically, Si- and C-terminated SiC substrates were heated in Mg vapor at 740°C for 1 minute. In Fig. 4, AFM images of these treated Si- and C-terminated SiC substrates are shown along with those for the pristine substrates as received from the vendor. For the "as received" material the Si-terminated surface (Fig. 4(a)) shows atomic steps with an RMS roughness of 0.2 nm whereas the C-terminated surface (Fig. 4(c)) is featureless with an RMS roughness of ~ 0.3 nm. Following the "Mg only" treatment the samples are essentially MgO/SiC structures due to oxidation of the air-exposed Mg layer. The roughness measured on Si-terminated substrate is much higher (RMS roughness ~ 4 nm) than that measured on the C-terminated substrate (RMS roughness ~ 0.4 nm).. They consistently show an oxygen-rich layer with Mg:O ratio of ~ 1:1.3 for both surface terminations and a small channeling effect indicative of imperfect or misaligned MgO crystallinity. The thickness of the $MgO_x$ layer is thicker on Si-SiC ($3.0 \pm 0.2$ nm as determined from the channeling spectrum) than on C-SiC: ($2.1 \pm 0.2$ nm). Note that they are both thicker than the $MgO_x$ layers detected at the $MgB_2$/SiC interface: $2.3 \pm 0.2$ nm on the Si face and $0.9 \pm 0.2$ nm on the C face, due to further oxidation upon air exposure.

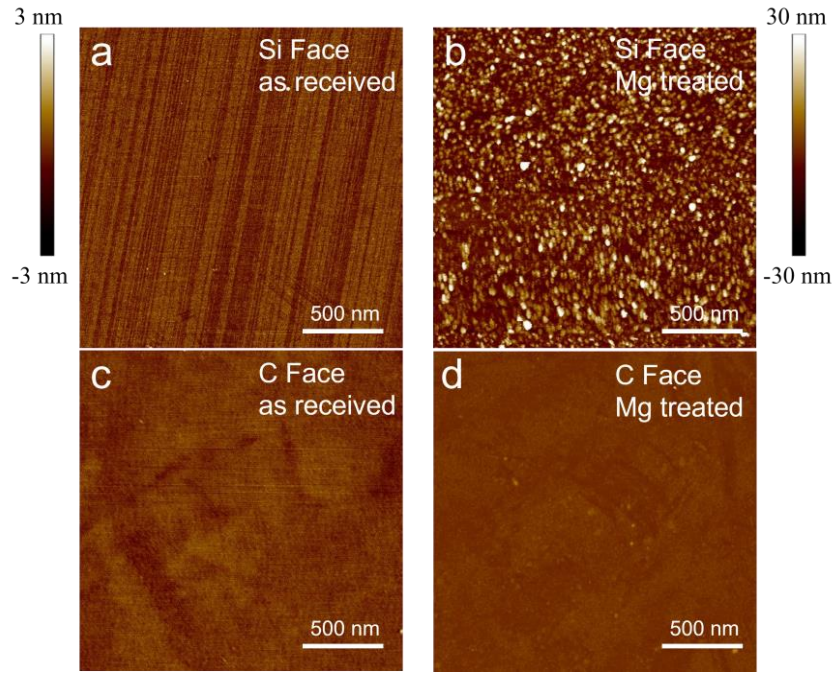

Figure 4. (a) Atomic Force Microscopy (AFM) of Si-SiC as received from the vendor, (b) AFM image of Si-SiC after annealing in Mg vapor at 740°C for one minute, (c) AFM scan for C-SiC as received from manufacture, (d) AFM image of C-termination of SiC after annealing in Mg vapor at 740°C for one minute.

### iii-b STEM Electron energy loss spectroscopy and nm elemental profiling

To further examine the interface between the $MgB_2$ films and the underlying C- and Si-terminated SiC substrates, cross-section samples were prepared and studied by STEM-EELS. As shown in the atomic resolution HAADF images in Figure 5, there is an intermediate layer between the top $MgB_2$ and the bottom SiC for both terminations. This interface layer is about 0.9 nm in both cases. However, while not shown here, the intermediate layer does not have constant thickness across the observed interface ranges, and it varies from 0.9 to 2.7 nm for the case of Si-SiC and from 0.9 to 1.8 nm for the case of C-SiC. In addition, the intermediate layer marked with dash lines between 0 nm to -1 nm in both cases often exhibits a periodic structure as shown in the figure. Even though the resolution of STEM image for the MgO layer is not ideal due to the combination of limited resolution of our instrument and the complex and imperfect thin structure, we can still see that the structure matches well with MgO as viewed from the (111) direction. The atomic arrangements of $MgB_2$, MgO, and SiC are overlaid in the figure as visual guides. Note that the image intensity scales with the atomic number due to the detector arrangement, and it is Mg and Si observed in the $MgB_2$ and SiC, respectively.

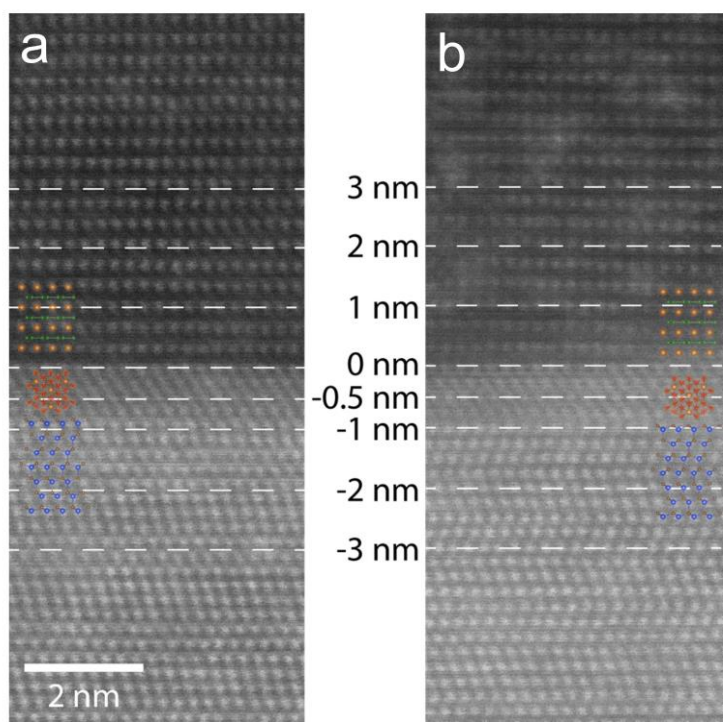

Figure 5. Cross-sectional HAADF STEM imaging of MgB$_2$ on (a) C-terminated SiC(0001) substrate (b) Si-terminated SiC(0001) substrate.

Next, we compare the relative chemical distributions of Mg, Si, B, C, and O along the eight data points acquired across the interface areas in Figure 6. It is found that oxygen is mostly confined in the intermediate layer. Si and C are found up to terminate near p the intermediate layer, as expected, whereas B is found on the thin film/surface side of the intermediate layer. Finally, Mg is found throughout the intermediate layer, consistent with a MgO interface.

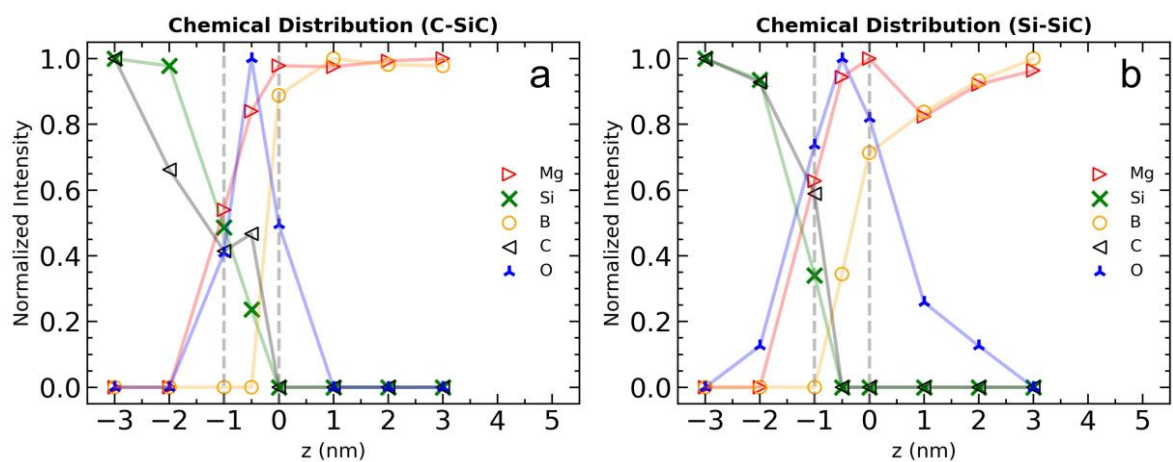

Figure 6. STEM-EELS data from the same cross section for (a) C-terminated SiC(0001) substrate (b) Si-terminated SiC(0001) substrate.

**iii-c) XPS analysis and interfacial chemistry**

More detailed information on interfacial chemistry is revealed by XPS. Figure 7 shows selected core levels spectra (Mg $1s$, Si $2p$, O $1s$ and C $1s$) measured on bare SiC substrates (bottom curves), Mg vapor treated SiC (middle curves), and 7 nm-thick $MgB_2$ grown on SiC (top curves), for both Si- and C-terminated surfaces. In all cases, the SiC substrate signal is detectable via the Si $2p$ and C $1s$ core levels. Chemical environments attributed to core levels features are indicated in the figure. On the bare SiC surface, exposed to air after preparation, the substrate Si $2p$ and C $1s$ core levels are found at binding energies of ~100 eV and ~282 eV, respectively, in good agreement with expected values [38]. The features in the O $1s$ and C $1s$ core levels spectra indicate carbohydrates adsorption, both the results of air exposure of the bare SiC substrates.

For the Mg vapor treated SiC samples, the O $1s$ level is split into two peaks. The lower binding energy component is attributed to MgO while the higher binding energy component is assigned to $Mg(OH)_x$ and Mg carbonates [39]. Of particular interest is that the SiC-related core level spectra show different binding energies for the two different SiC surface terminations: the binding energies of both the C $1s$ and Si $2p$ core levels are ~1 eV lower for the C face than for the Si face. For $MgB_2$ films on SiC substrate, exposure to air causes surface oxidation as well as water and carbohydrates adsorption. As a result, their O $1s$ and Mg $1s$ spectra are affected by both the top surface alteration of $MgB_2$ and the interfacial $MgO_x$ layer, and separating these contributions is not straightforward. However, we again observe a shift of the Si $2p$ core level to lower binding energy by ~1 eV in the sample on the C face as compared to the Si face (a similar energy shift is present for C 1s, but less visible due to the SiC signal attenuation through the $MgB_2$ layer). This suggests that in the cases of both Mg vapor treated SiC, for which the stack is effectively MgO/SiC, and for $MgB_2$ films on SiC, for which the stack is likely $MgB_2$/MgO/SiC, there is a similar energy alignment related to the MgO/SiC interface that is dependent on the SiC surface termination. Similar behavior has been reported in the case of intentionally grown epitaxial MgO films on SiC surfaces [40,41]: binding energy offsets of the order of eVs, have been measured for different MgO/SiC interfaces, supported by electronic structure calculations of atomically different interfaces.

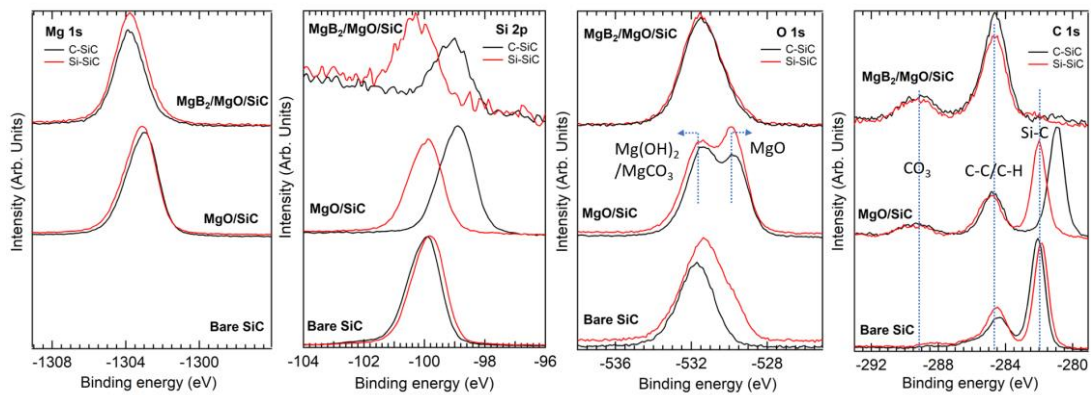

Figure 7. XPS spectrum of Mg $1s$, Si $2p$, O $1s$ and C $1s$ for bare SiC, Mg vapor treated SiC, and thin $MgB_2$ film on SiC for both C and Si terminations.

## IV. Discussion

In the following, we consider specific aspects of these analyses and how they might influence the film morphology.

- Effect of native oxide

It is clear from the different interface analysis results above, that a buried MgO layer exists at the interface of the $MgB_2$ film and the SiC substrate. It is likely that this oxide is the result of the reaction of Mg with the "native oxide" that exists on the SiC surface. The reaction of Mg with $SiO_2$ has been reported by a number of authors suggesting that the reaction of Mg with $SiO_2$ can result in formation of MgO and possibly Mg silicates [42, 43].

"Native oxide" as a thin film is not necessarily well defined, as the resulting oxide thickness, usually only 1-2 nm at most, is a result of environmental variables, time and crystal face. In recent work using contact angle measurements Park et al [44] showed that the native oxide growth is greater on the Si face than on the carbon face, consistent with the reports in Table 2 of a Si face MgO layer of 2.3 nm compared to a C face of 0.9 nm. For calibration, we note that 1 nm of $SiO_2$ corresponds to $5.3\times10^{15}$ /$cm^2$ of oxygen, consistent with the existence of very thin native oxides yielding nanometer MgO. Furthermore, Nagai et al [45] characterized the roughness in very thin oxides on SiC as a function of crystal face. In this work, the authors show that the RMS roughness is proportional to oxide thickness and the rate of increase of roughness with film thickness is the same for the two crystal faces. Therefore, the Si face oxide roughness is greater than the C face. These reports allow some mechanistic conclusions as follows. Native oxide thickness on the C face is less than that of the Si face [44], consistent with the reports in Table 2. The roughness of very thin oxides is proportional to oxide thickness [45]. In short, oxide roughness is proportional to oxide thickness, this oxide roughness is transferred to a MgO layer and then reflected in the overlayer $MgB_2$ film uniformity. Since the C-face oxide is substantially less than the Si face the net roughness is reduced for the C-face resulting in a more uniform thin film.

Relevant to that point, it is interesting to note that there are reports of growth of MgO on SiC by MBE for MOS systems [46, 47], a good lattice match, and there are reports of the growth of $MgB_2$ on MgO [48], also a reasonable lattice match resulting in high quality $MgB_2$ films. Therefore, an $MgB_2$/MgO/SiC epitaxial structure may be realized. Nevertheless the properties of the resulting film but may depend on the SiC termination: if the starting (oxidized) surface of the C face of SiC is less rough than the Si face, a smoother MgO/$MgB_2$ structure is expected.

Finally, among the interesting remaining questions is the "necessity" for magnesium-based oxide layer to achieve higher quality epitaxy and crystallinity. We note this point was explicitly raised in the MBE work of Laloe et al [49], for $MgB_2$ on Si where a Mg starting layer was explicitly added to enhance growth. Possibly the "native oxide" on the Si face is just the correct amount to achieve a high-quality epitaxial film. In that regard, we note that some preliminary experiments on HF treated SiC (presumably minimal oxide) in our laboratory did not produce quality films.

## V. Conclusion

The growth of ultrathin $MgB_2$ films on the different surface terminations of SiC has been studied, seeking the conditions for optimum superconducting properties and film uniformity. It has been shown that the best conditions are associated with 10 sccm of 5% diborane gas flow rate on the C terminated face of SiC for our specific system. A significant difference has been identified between growth on the Si face and the C face, with the latter producing higher quality films. This difference has been explored by various interface probes.

RBS/channeling measurements indicated that the samples consisted of $MgB_2$/MgO/SiC stacks, in which the thickness $MgO_x$ layer was SiC surface termination dependent: is $2.3 \pm 0.2$ nm on the Si face and $0.9 \pm 0.2$ nm on the C face. High resolution EELS and TEM confirmed this structural difference, indicating that different interfacial constituents on the two surfaces may control the final morphology. XPS analysis indicated a similar energy band offset at the MgO/SiC interfaces, for both $MgB_2$ films grown on SiC and for MgO films on SiC, but highly dependent on the SiC surface termination. High energy, grazing exit angle XPS confirmed the presence of a buried, thin $MgO_x$ layer at the $MgB_2$/SiC interface.

This MgO layer in turn may govern the $MgB_2$ film quality: a thicker and rougher $MgO_x$ layer on the Si face of the SiC substrate is the cause of the rougher ultrathin $MgB_2$ films as compared to the films on the C face of SiC. The achievement of such high-quality superconducting films, and the knowledge of the parameters that control their growth, maybe a precursor to new devices and device configurations employing their unique electronic properties.

## Acknowledgment

Research at Temple University was supported by NASA's Astrophysics Research and Analysis Program through a contract from JPL (Contract No. 1632463). Research at Rutgers was supported by Laboratory for Surface Modification (LSM). A.X.G., R.K.S., and J.R.P. acknowledge support from the DOE, Office of Science, Office of Basic Energy Sciences, Materials Sciences, and Engineering Division under Award No. DE-SC0019297. The electrostatic photoelectron analyzer for the lab-based HAXPES measurements at Temple University was acquired through an Army Research Office DURIP grant (Grant No. W911NF-18-1-0251).

# Supplementary Information

**Ultra-thin Epitaxial MgB$_2$ on SiC: Substrate Surface Polarity Dependent Properties**

Weibing Yang[1†*], Leila Kasaei[2†], Hussein Hijazi[2], Sylvie Rangan[2], Yao-wen Yeh[2], Raj K Sah[1], Jay R Paudel[1], Ke Chen[1], Alexander X. Gray[1], Philip Batson[2], Leonard C. Feldman[2], and Xiaoxing Xi[1]

[1]*Department of Physics, Temple University, Philadelphia, PA 19122, USA*

[2]*Department of Physics and Astronomy, Rutgers University, New Brunswick, NJ 08901, USA*

*Corresponding author

# 1. Hybrid Physical Chemical Vapor Deposition of epitaxial MgB$_2$ thin films.

The HPCVD reactor was first purged with ultra-high purity H$_2$ gas. The substrate and Mg pieces placed nearby on the same heater were then heated to 740°C in the H$_2$ ambient when Mg began to evaporate. After about 10 seconds, a mixture of 5% B$_2$H$_6$ in H$_2$ was introduced into the reactor to initiate growth. The film deposition rate can be controlled by adjusting the flow rates of the diborane gas mixture. In this work, a flow rate of 10 sccm was used, the result of parameter optimization for ultrathin MgB$_2$ film with the best film quality. The corresponding deposition rate was ~ 0.23 nm/s, determined by a linear fitting of the thicknesses-flow rate data from a series of calibration runs. The thickness of MgB$_2$ ultrathin films was then controlled by the deposition time. The deposition temperature was 740 °C.

In addition to figure 3 in the main text, we present in figure S1 the residual resistivity ($\rho_0$) of MgB$_2$ films as a function of thickness for many more samples, measured in unpatterned as-grown films, to demonstrate the reproducibility of our deposition process. The additional data points are derived from the measurements of MgB$_2$ films grown using 5, 10, and 20 sccm gas mixture of 5% B$_2$H$_6$ in H$_2$. It is clear that the flow rate of B$_2$H$_6$ gas mixture does not change the trend of residual resistivity as a function of film thickness. The residual resistivity is about the same for thick films on the Si and C faces, but it is much smaller on the C face than on the Si face when the film is only a few nanometers, indicating much better gain connectivity of ultrathin MgB$_2$ on the C face.

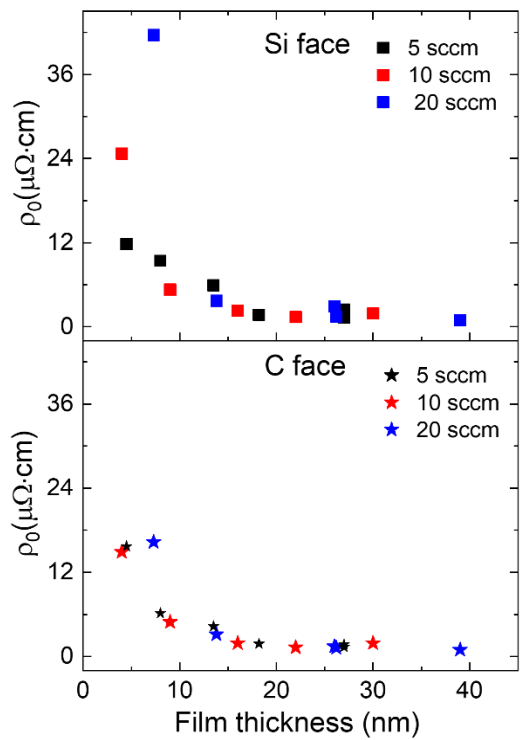

Figure S4. Residual resistivity as a function of MgB$_2$ film thickness on the Si and C faces. The MgB$_2$ films were deposited using 5, 10 and 20 sccm gas mixture of 5% B$_2$H$_6$ in hydrogen.

## 2. Comparison of Rutherford Back Scattering on MgB$_2$ thin films grown on two terminations of SiC.

Figure S2(a) shows a SIMNRA simulation of a random (non-channeling) RBS spectrum for an ideal 80 nm MgB$_2$ film grown epitaxially on SiC substrate. The thickness and scattering geometry was chosen such that signals from the film surface and film/substrate interface are clearly distinguishable. Figures S2(b) and S2(c) show RBS channeling and random spectra for two 80 nm MgB$_2$ films on Si- and C-terminated SiC substrates, respectively. The experimental random RBS spectra agree well with the SIMNRA fitting shown in Fig. S2(a). The minimum yield $\chi_{min}$ (the ratio between the channeling and random yields), evaluated using a width of 3 channels at around channel number 665, with contributions from both the MgB$_2$ and the underlying SiC, was ~5%, indicating excellent crystallinity in MgB$_2$ films on both the Si and C faces.

The strong channeling effect allows us to probe the properties of the MgB$_2$/SiC interface as an alignment discontinuity at the interface inevitably leads to a weaker channeling effect. In Figs. S2(b) and S2(c), the channeling yields are multiplied by 5x for clarity. For both SiC terminations, there is an interfacial layer between the MgB$_2$ film and the SiC substrate containing both magnesium and oxygen, marked as "Mg interface" and "O interface", respectively. We identify this interfacial layer as MgO$_x$. The degraded channeling effect in the interfacial layer may be the result of partially crystalline MgO$_x$ and/or lattice mismatch between MgB$_2$ and MgO$_x$.

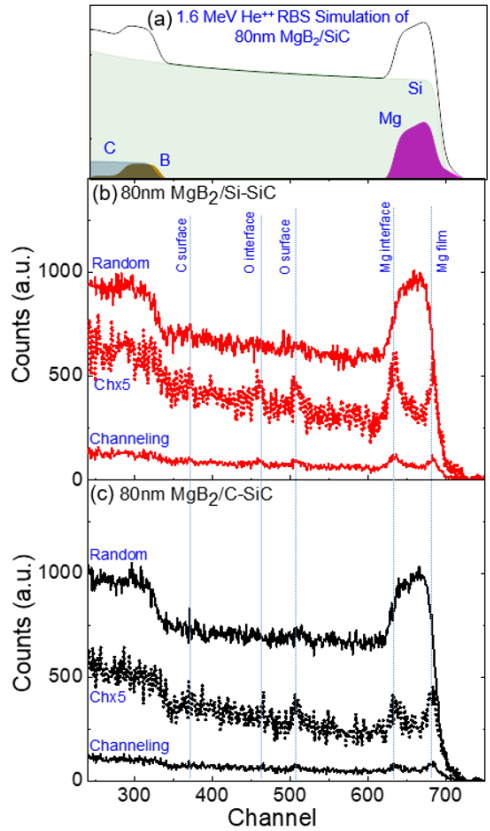

Figure S2. (a) Simulated random RBS spectrum for an ideal 80 nm MgB$_2$ films grown on SiC. (b) and (c) RBS channeling and random spectra for 80 nm MgB$_2$ films on Si- and C-terminated SiC(0001) substrates, respectively. The channeling yields multiplied by 5x are also shown for clarity.

## 3. Angle Resolved Hard X-ray Photoelectron Spectroscopy on MgB$_2$ thin films on two terminations of SiC substrates

To investigate the chemical bonding and the depth profile of the Mg species associated with the pristine MgB$_2$ and the oxidized film, we utilized angle-resolved bulk-sensitive hard x-ray photoelectron spectroscopy (HAXPES) [1]. The measurements were carried out using a lab-based HAXPES system equipped with a monochromatized Cr Kα x-ray source with the photon energy of 5.4 keV and a wide acceptance angle hemispherical electrostatic analyzer ScientaOmicron EW4000.

We measured two MgB$_2$ films, nominally 4 nm thick, deposited simultaneously and side-by-side on the C- and Si-terminated SiC substrates. At the photon energy of 5.4 keV, the inelastic mean-free path (IMFP) for the Mg 1$s$ photoelectrons ($E_{kin}$ = 4.1 keV) in MgB$_2$ is estimated to be approximately 7 nm [2], which ensures that the entire film and the interface with the substrate are being probed. Measurements were carried out at four different photoelectron take-off angles, 4°, 30°, 45°, and 60°, facilitating different average probing depths varying from approximately 7 nm (at 4°) to approximately 3.5 nm at 60°, making the latter more surface sensitive and less sensitive to the buried SiC/MgB$_2$ interface [3].

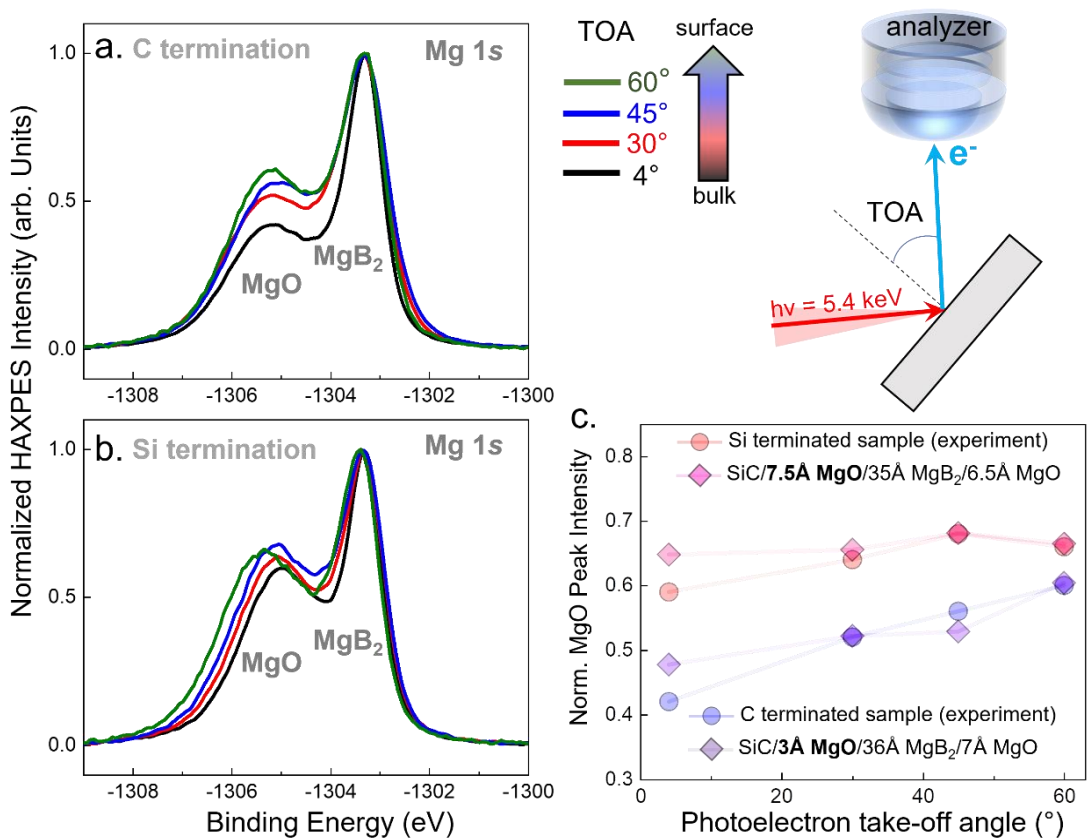

Figure S3. (a) Mg 1s core-level peak measured on the MgB$_2$ film deposited on the C-terminated SiC substrate at four photoelectron take-off angles in the experimental geometry shown on the right. (b) Same angle-resolved measurement carried out on the MgB$_2$ film deposited on the Si-terminated SiC substrate. (c) Peak intensities of the higher-binding-energy MgO peak component normalized to the height of the main MgB$_2$ peak (circular markers) and the corresponding simulated intensities (diamond-shaped markers) for the best-fit film structures shown in the figure legend.

The results of the measurements for the films on the C-terminated and Si-terminated SiC substrates are presented in Figures S3(a) and S3(b), respectively. The most intense peak at the binding energy of 1303.3 eV in both plots corresponds to the Mg 1$s$ core-level photoemission originating from the MgB$_2$ film [4]. The higher-binding-energy feature at approximately 1305.2 eV corresponds to the chemically shifted state originating, most likely, from the Mg oxide species, such as MgO [5]. Normalization of the photoemission intensities to the maximum of the main MgB$_2$ peak reveals two different trends in the angle-dependent evolutions of the photoemission intensities of the higher-binding-energy (MgO) component.

For the film on the C-terminated SiC substrate [Fig. S3(a)] the intensity of the MgO component increases with increasing surface sensitivity, as shown using blue circular markers in Figure S3(c). Such a trend generally corresponds to the angle-resolved measurement of the surface oxide species [6].

Conversely, for the film on the Si-terminated substrate [Fig. S3(b)] two major differences are observed. First, the average intensity of the MgO component is increased relative to the main MgB$_2$ peak, suggesting a higher oxide content in the sample. Secondly, the angle-dependent evolution of the MgO component exhibits a flatter trend, with the most surface-sensitive measurement at 60° exhibiting a slight decrease in intensity. Such a trend, shown using red circular markers in Figure S3(c), suggests the presence of a buried Mg oxide layer at a depth that is larger compared to the probing depth at the take-off angle of 60° (3.5 nm).

In order to confirm the presence of the buried Mg oxide layer in the sample on the Si-terminated SiC substrate, we carried out angle-resolved simulations using the SESSA simulation package, which quantitatively predicts photoemission peak intensities by taking into account relevant parameters such as IMFP, elastic-scattering cross-sections, photoionization asymmetry parameters, and the photoelectron take-off angles [7]. The best-fit results for both samples are shown using diamond-shaped markers in Figure S3(c). For both samples, the thickness of the surface oxide is predicted to be approximately 0.65 – 0.7 nm. The thickness of the pristine $MgB_2$ layer (3.5-3.6 nm) is close to the target thickness of 4 nm. The main difference between the two samples is the extracted thickness of the buried interfacial oxide, which is mainly responsible for the difference in the two observed intensity trends. Specifically, the best-fit thickness of the buried Mg oxide at the interface between $MgB_2$ and the Si-terminated substrate is 0.75 nm, while the same value for the C-terminated substrate is only 0.3 nm. The presence of an interfacial oxide was necessary for obtaining reasonable fits for both datasets. As explained in the discussion the interfacial oxide on these samples is determined by the "native oxide" on the SiC surface. Such native oxides are not well defined and depend on numerous environmental factors. Nevertheless, in agreement with the trend reported in Table 1, the MgO interfacial layer is greater on the Si face than on the C face.

Slight discrepancies between the experimental and simulated intensities are observed for both samples at the lowest take-off angle (4°) due to the limitations of the experimental geometry, which features a fixed 90° angle between the x-ray incidence direction and the analyzer orientation. In this geometry, the x-ray incidence angle is so grazing (4°) that the effects of the x-ray beam cone-angle (18°), as well as the total external reflection, may play a significant role.